\pdfoutput=1

\documentclass{sigchi}

\toappear{}

\usepackage{balance}  
\usepackage{graphicx} 
\usepackage{times}    
\usepackage{url}      
\usepackage{booktabs}
\usepackage{tabularx} 
\usepackage[sort,compress]{cite}

\usepackage[inline=true,margin=false]{fixme}
\newcommand{\edit}{}

\makeatletter
\def\url@leostyle{%
  \@ifundefined{selectfont}{\def\UrlFont{\sf}}{\def\UrlFont{\small\bf\ttfamily}}}
\makeatother
\def\pprw{8.5in}
\def\pprh{11in}

\usepackage[pdftex]{hyperref}
\hypersetup{
pdftitle={Unsupervised Motion Artifact Detection in Wrist-Measured Electrodermal Activity Data},
pdfauthor={Yuning Zhang, Maysam Haghdan, and Kevin S. Xu},
pdfkeywords={wearable; artifact detection; electrodermal activity; skin conductance; anomaly detection},
bookmarksnumbered,
pdfstartview={FitH},
colorlinks,
citecolor=black,
filecolor=black,
linkcolor=black,
urlcolor=black,
breaklinks=true,
}

\newcommand\tabhead[1]{\small\textbf{#1}}

\renewcommand{\th}{\tabhead}

\begin{document}

\title{Unsupervised Motion Artifact Detection in Wrist-Measured Electrodermal Activity Data}


\numberofauthors{1}
\author{
  \alignauthor Yuning Zhang, Maysam Haghdan, and Kevin S.~Xu\\
    \affaddr{EECS Department, University of Toledo, 
    Toledo, OH, USA}\\
    \email{\{Yuning.Zhang, Maysam.Haghdan\}@rockets.utoledo.edu, Kevin.Xu@UToledo.edu}\\
}

\maketitle

\begin{abstract}
One of the main benefits of a wrist-worn computer is its ability to collect a variety of physiological data in a minimally intrusive manner. 
Among these data, electrodermal activity (EDA) is readily collected and provides a window into a person's emotional and sympathetic responses. 
EDA data collected using a wearable wristband are easily influenced by motion artifacts (MAs) that may significantly distort the data and degrade the quality of analyses performed on the data if not identified and removed. 
Prior work has demonstrated that MAs can be successfully detected using supervised machine learning algorithms on a small data set collected in a lab setting.  In this paper, we demonstrate that {\edit unsupervised learning algorithms perform competitively with supervised algorithms} for detecting MAs on EDA data collected in both a lab-based setting and a real-world setting comprising about 23 hours of data. We also find, somewhat surprisingly, that {\edit incorporating accelerometer data as well as EDA improves detection accuracy only slightly for supervised algorithms and significantly degrades the accuracy of unsupervised algorithms}.
\end{abstract}

\section{Introduction}

With the increasing popularity of wearable computers on the wrist, including fitness bands and smart watches, there has been tremendous interest on analyzing data collected from these wearables, particularly physiological data. Electrodermal activity (EDA) is readily measured by a wearable wristband and reflects the emotional and sympathetic responses of a person \cite{boucsein2013electrodermal}. 
EDA has been used in many applications including {\edit stress detection \cite{hernandez2011call}, measuring human engagement \cite{Hernandez2014},} and classifying autonomic nervous system activity \cite{Natarajan2016}. 

\begin{figure}[t]
\centering
\includegraphics[width=3.4in]{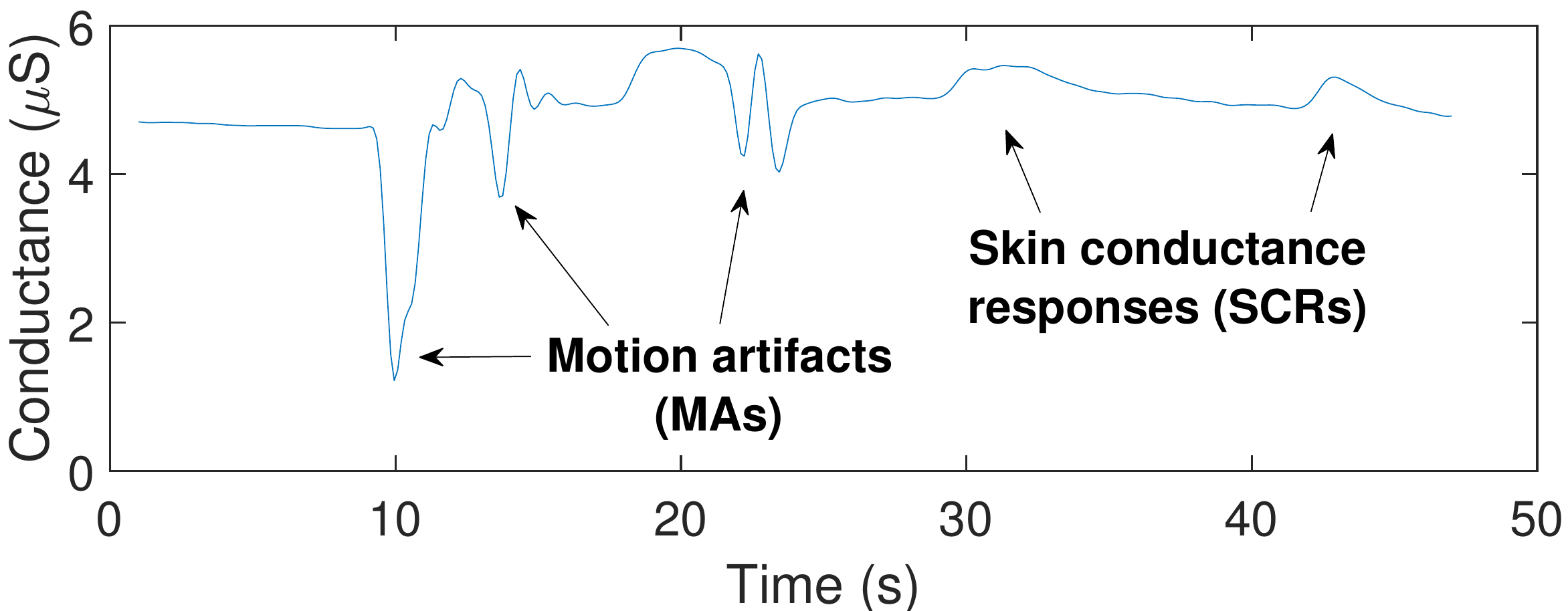}
\caption{Examples of SCRs compared to MAs. A sudden increase in SC may be indicative of the start of an SCR or an MA.}
\label{fig:MAsSCRs}
\end{figure}

EDA is commonly measured via the skin conductance (SC). When a person is under stress or at a high level of emotion, the sympathetic nervous system is activated and causes the person to sweat, increasing the SC in a series of skin conductance responses (SCRs) where the SC rapidly increases then gradually decays. 
EDA data have traditionally been collected using stationary equipment in a laboratory setting; however, recent wearable wristbands, such as the Affectiva Q sensor \cite{affectivaWhite}, offer the ability to non-invasively measure EDA in real-world environments. 
One of the main challenges when analyzing EDA data collected from such wearables is the presence of motion artifacts (MAs) in the data. 
Such artifacts may result from {\edit changes in the amount of pressure on the sensor} as well as movements or rotations of the wrist that affect the amount of contact between the electrodes and the skin. 
Some examples of SCRs and MAs are shown in Figure \ref{fig:MAsSCRs}. 

MAs in EDA data can significantly degrade the quality of analyses performed on the data; thus, MAs should first be suppressed or detected.
MA suppression attempts to clean the portions of data with MAs by passing {\edit the entire SC time series} through some type of smoothing filter \cite{chen2015wavelet,hernandez2011call,Hernandez2014}. The main downside to MA suppression is that it distorts the SC even outside of the portions with MAs, including the informative SCRs. 
MA detection, on the other hand, aims to identify portions of the data with MAs so they can be removed from further analysis. Taylor et al.~\cite{Taylor2015} formulated MA detection as a supervised machine learning problem and demonstrated that supervised learning algorithms can automatically detect  MAs on a small EDA data set collected in a lab environment. The downside to supervised algorithms is that they require lots of labeled data to train, which requires significant human effort. 
Additionally, to the best of our knowledge, prior work on suppressing and detecting MAs in EDA has not taken advantage of data collected from an accelerometer, which is also typically present on wearable wristbands. 

In this paper, we apply {\edit 8} different machine learning algorithms, {\edit 5} supervised and {\edit 3} unsupervised, to two EDA data sets comprising about 23 hours of data in both lab and real-world settings, to automatically detect MAs in EDA. We also evaluate the usefulness of accelerometer data for improving MA detection. 
Our main findings are as follows:
\begin{itemize}
\item {\edit The accuracy of unsupervised learning algorithms is competitive with that of the supervised algorithms for both in-sample (within a particular data set) and out-of-sample prediction (when training and testing on different data sets).}
\item Inclusion of the accelerometer data {\edit only slightly improves} the accuracy of supervised learning algorithms and significantly degrades the accuracy of unsupervised algorithms.
\end{itemize}
The comparatively strong performance of unsupervised algorithms is very promising because they potentially enable MA detection on a large scale without significant human effort in labeling training data, which addresses a significant problem in analyzing EDA data collected using wearables.

\section{Data Description}
We use two publicly available data sets with EDA and 3-axis accelerometer data, both collected using an Affectiva Q sensor \cite{affectivaWhite} worn on the wrist, totaling about 23 hours of data.

\subsection{UT Dallas Stress (UTD) Data}

This data set was collected at the University of Texas at Dallas \cite{Birjandtalab2016}. A total of 20 college students (14 males and 6 females) were asked to perform a sequence of tasks subjecting them to three types of stress: physical stress (standing, walking, and jogging), cognitive stress (mental arithmetic and the Stroop test), and emotional stress (watching a horror movie clip). Each task was performed for 5 minutes, and tasks were separated by 5 minute relaxation periods. Altogether, about 13 hours worth of data were collected.
Over all 20 subjects, {\edit 3.8\% of the data was determined by three human experts to contain MAs (see \nameref{sec:Labeling} section for details). On the low end, three subjects' data contained no MAs; on the high end, one subject's data contained 14\% MAs.}

\subsection{Alan Walks Wales (AWW) Data}

This data set was collected by Alan Dix while he walked around Wales from mid-April to July 2013 \cite{Dix2017}. He collected 64 days of data and also wore a GPS and kept a diary of his activities. {\edit We extracted segments of data over 10 different days resulting in 10 hours of data in total. We split the segments into two categories of activities: walking and resting.} The walking data contain 5 hours of data collected as Alan was walking or hiking, and the resting data contain 5 hours of data collected when he was resting, eating, reading, or interacting with others. The reason we divided the data in this way is that the walking data contain more physical movements, which in turn have more MAs, while the resting data contain less physical movements (and less MAs) but more cognitive and emotional activities.
{\edit 33\% and 15\% of the walking and resting data, respectively, were determined to contain MAs, making it a more challenging data set than UTD.}

\section{Methods}
\subsection{Feature Construction}
{\edit Following the analysis in \cite{Taylor2015}, we divide the data} into 5-second time windows.  For each time window, we construct statistical features on both the EDA and simultaneously collected 3-axis accelerometer data. 
{\edit For the EDA data, we consider 6 different signals: the SC amplitude, its first and second derivatives, and the coefficients of a Discrete Wavelet Transform (DWT) with the Haar wavelet applied to the SC at 3 different time scales: 4 Hz, 2 Hz, and 1 Hz. 
Wavelet transforms are able to capture both frequency and time information, and the Haar wavelet is excellent at detecting sudden changes in signals, which frequently occur during MAs. 
The 6 signals we consider were found to be informative for MA detection in EDA by Taylor et al.~\cite{Taylor2015}. 
For each of the 6 signals, we construct 4 statistical features: the mean, standard deviation, maximum, and minimum over the 5-second windows, resulting in 24 total EDA features.

We construct the same set of features as for EDA on each of the 3 axes of accelerometer data, as well as on the acceleration magnitude (root-mean-square). This results in 24 features for each of the 3 axes and 24 features for the magnitude for a total of 96 accelerometer features.
We arrived at this set of features after examining a significant amount of prior work on classification using EDA \cite{hernandez2011call,Hernandez2014,Taylor2015,Natarajan2016} and accelerometer \cite{Lara2013,Hammerla2015} data. The final selection of features is admittedly somewhat ad-hoc; however, we believe it is a fair representation of commonly used features in the literature.}

\subsection{Machine Learning Algorithms}
{\edit We examine 5 supervised algorithms: support vector machines (SVMs), k-nearest neighbor (kNN) classifiers, random forests, logistic regression, and multi-layer Perceptrons (MLPs). We also examine unsupervised variants of the first 3 supervised algorithms: 1-class SVMs, kNN distances, and isolation forests, respectively.} The kNN distance algorithm differs from the kNN classifier in that it uses the distance between a test sample and its k-th nearest training sample as its test statistic, whereas the kNN classifier performs a majority vote over the labels of the k nearest training samples. 

The supervised algorithms are used for binary classification, where the two classes are MA and clean. The unsupervised algorithms are used for anomaly detection, where it is assumed that the training data consists of mostly clean data. We interpret the time windows predicted by the unsupervised algorithms as anomalies to be predicted MAs.

{\edit We optimize the parameters of each algorithm by using a grid search with exponential grid and retain the parameters with the highest cross-validation accuracy (see \nameref{sec:Results} section for details).} 
We use the Gaussian kernel for the SVM, which was found to be most accurate in \cite{Taylor2015}, {\edit and ReLU activations for the MLPs with up to 2 hidden layers. 
For the unsupervised algorithms, we also use the grid search cross-validation approach to select parameters in order to provide a fair comparison to the supervised algorithms. 
Since this approach is likely not possible in practice with unlabeled data, we also experiment with different choices of parameters to determine the sensitivity of results to the parameter choices.}

\section{Experiment Set-up}
The experiments involve evaluating the predictions of the machine learning algorithms compared to hand-labeled MAs by {\edit three} EDA experts. 
Code to reproduce the experiments is available at 
\url{https://github.com/IdeasLabUT/EDA-Artifact-Detection}.

\begin{table*}
\centering
\caption{In-sample prediction AUC using {\edit leave-one-subject-out CV. The top five algorithms are supervised, while the bottom three are unsupervised.}}
\label{tab:CvAuc}
\addtolength{\tabcolsep}{-1pt}
\begin{tabular}{cccccccccc}
\hline
                               & \multicolumn{3}{c}{\th{Alan Walks Wales resting data}} & \multicolumn{3}{c}{\th{Alan Walks Wales walking data}} & \multicolumn{3}{c}{\th{UT Dallas data}}           \\
\th{Algorithm}                 & \th{All features} & \th{ACC only} & \th{EDA only}      & \th{All features} & \th{ACC only} & \th{EDA only}      & \th{All features} & \th{ACC only} & \th{EDA only} \\
\hline
Logistic regression            & 0.843             & 0.714         & 0.775              & 0.807             & 0.649         & 0.796              & {\bf 0.941}       & 0.852         & 0.935         \\
{\edit Multi-layer Perceptron} & 0.683             & 0.539         & 0.696              & 0.788             & 0.663         & 0.777              & 0.928             & 0.842         & 0.928         \\
SVM                            & 0.689             & 0.582         & 0.688              & 0.798             & 0.684         & 0.782              & 0.913             & 0.852         & 0.898         \\
kNN classification             & 0.674             & 0.582         & 0.738              & 0.740             & 0.641         & 0.776              & 0.846             & 0.832         & 0.870         \\
{\edit Random forest}          & 0.747             & 0.583         & 0.712              & 0.815             & 0.671         & 0.796              & 0.935             & 0.852         & 0.937         \\
\hline
1-class SVM                    & 0.844             & 0.763         & 0.850              & 0.768             & 0.683         & 0.760              & 0.859             & 0.862         & 0.900         \\
kNN distance                   & 0.807             & 0.723         & {\bf 0.898}        & 0.774             & 0.705         & {\bf 0.847}        & 0.911             & 0.875         & 0.930         \\
{\edit Isolation forest}       & 0.804             & 0.711         & 0.885              & 0.693             & 0.619         & 0.735              & 0.909             & 0.878         & 0.900         \\
\hline
\end{tabular}
\end{table*}

\subsection{Expert Labeling}
\label{sec:Labeling}
We have {\edit three} EDA experts hand label each 5-second time window as clean or containing an MA using the EDA Explorer software \cite{EdaExplorer,Taylor2015}. {\edit All experts used a common set of criteria} to define an MA in the SC:
a peak that does not have an exponential decay, except when two peaks are very close to each other in a short time period so that the decay of the first peak is interrupted by the second peak;  
a sudden change in SC correlated with motion; or  
a sudden drop of more than 0.1 $\mu$S in SC. 
{\edit The first two criteria were used in \cite{Taylor2015}; we added the third criterion based on the physiology of EDA. 
SC can increase suddenly due to sweat glands releasing sweat, but 
there is no physiological mechanism for SC to decrease suddenly \cite{boucsein2013electrodermal}.}

{\edit The labels from all three experts were in agreement for 95\% and 87\% of the time windows for the UTD and AWW data, respectively. 
The only two possibilities for disagreement are two experts labeling as MA and one as clean or vice-versa. 
When there was disagreement, 2 MA/1 clean occurred 38\% and 44\% of the time in the UTD and AWW data, respectively.}

\subsection{In-sample Prediction}
{\edit For the UTD data, we evaluate the in-sample prediction accuracy for each learning algorithm using a leave-one-subject-out cross-validation (CV) approach, which was found to be preferable to k-fold CV for time series data due to dependence of time windows \cite{Hammerla2015}. 
For the AWW data, we have only one single subject, but we have 10 hours of labeled time segments spaced out across 10 different days in the data trace, so we adopt a leave-one-segment-out approach.}
For both data sets, we use the Area Under the Curve (AUC) of the Receiver Operator Characteristic as the accuracy metric. 

We split the Alan Walks Wales data into two separate data sets containing only resting data and only walking data prior to performing the CV to evaluate the prediction accuracy for both categories of activities.
To test the value of the accelerometer data, we test the learning algorithms on three different feature sets: ACC only, containing only the {\edit 96} features constructed from the accelerometer; EDA only, containing the {\edit 24} features constructed from EDA; and all {\edit 120} features. 

\subsection{Out-of-sample Prediction}
For out-of-sample prediction, we train each learning algorithm on all of the time windows in one of the two data sets, then test the algorithm's predictions on all of the time windows in the other data set. 
In this experiment, we treat the Alan Walks Wales data as a single data set so that the two data sets are of roughly the same size. 
This should be a tougher prediction task than in-sample prediction because the two data sets contain different activities and were collected in very different settings with different test subjects.
{\edit For all 8 algorithms, we use the same parameter tuning method on the training data set as in the in-sample prediction task.}

\section{Results}
\label{sec:Results}

\subsection{In-sample Prediction}

{\edit The results of the in-sample prediction task are shown in Table \ref{tab:CvAuc}.} 
The first observation from all three data sets is that using only the accelerometer-derived features (ACC only) provides a significantly worse predictor than the other two feature sets. 
Additionally, in all three data sets, we do not observe a {\edit significant} benefit in using all of the features rather than just the EDA-derived features (EDA only). 
The inclusion of accelerometer-derived features appears to have minimal effect on the supervised learning algorithms, {\edit which only improve by 0.4\% on average, and the AUCs of the unsupervised algorithms actually decrease by 4.3\% on average.} 
Additionally, when comparing the AUCs of the supervised and unsupervised algorithms, {\edit we notice that the unsupervised algorithms (on EDA features only) perform very competitively with and sometimes even better than the supervised algorithms}. 
We expand on these points in the \nameref{sec:Discussion} section.

\subsection{Out-of-sample Prediction}

\begin{table}
\centering
\caption{Out-of-sample prediction AUC with EDA only features.}
\label{tab:OutSampleAuc}
\begin{tabular}{ccc}
\hline
                               & \multicolumn{2}{c}{\th{Train/Test Set}} \\
\th{Algorithm}                 & \th{AWW/UTD }            & \th{UTD/AWW} \\
\hline
Logistic regression            & 0.943            		  & 0.846        \\
{\edit Multi-layer Perceptron} & 0.943                    & {\bf 0.859}  \\
SVM                            & 0.944                    & 0.822        \\
kNN classification             & {\bf 0.946}              & 0.827        \\
{\edit Random forest }         & 0.940					  & 0.843		 \\
\hline
1-class SVM                    & 0.891                    & 0.847        \\
kNN distance                   & 0.913                    & 0.854        \\
{\edit Isolation forest }      & 0.911					  & 0.774		 \\
\hline
\end{tabular}
\end{table}

{\edit The results of the out-of-sample prediction task are shown in Table \ref{tab:OutSampleAuc}.} 
In this task, using {\edit EDA only features resulted in better performance in both data sets} than using the other feature sets, so we show results with EDA only features. 
Notice that the results are worse for all algorithms when training on the lab-based UTD data and testing on the real-world AWW data, as one might expect when attempting to generalize from data collected in {\edit a lab setting.}
Notice also that the unsupervised algorithms are again {\edit competitive with the supervised.

In practice, choosing parameters for the unsupervised algorithms is very difficult without labeled data, in which case CV is not possible.} 
We find that the kNN distance algorithm does not appear to be very sensitive to the choice of the number of neighbors. {\edit For example, by sweeping the number of neighbors from 1 to 30, the AUC varies by at most 0.01} when training on the AWW data and testing on the UTD data. {\edit Isolation forests are slightly more sensitive, with AUC varying by up to 0.05 for different parameter choices, and 1-class SVMs are the most sensitive with AUC varying by up to 0.2.}

\section{Discussion}
\label{sec:Discussion}

\begin{figure}[t]
\centering
\includegraphics[width=1.6in]{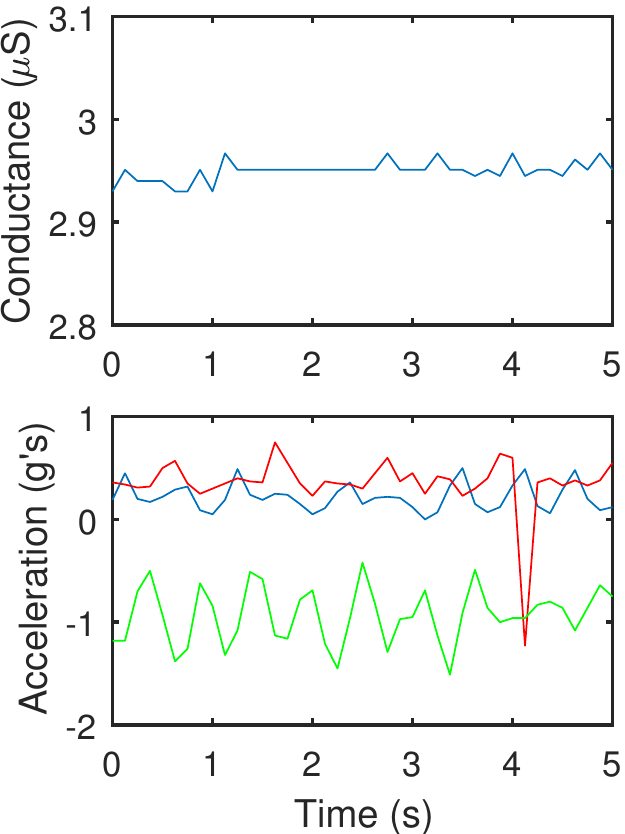}
\quad
\includegraphics[width=1.6in]{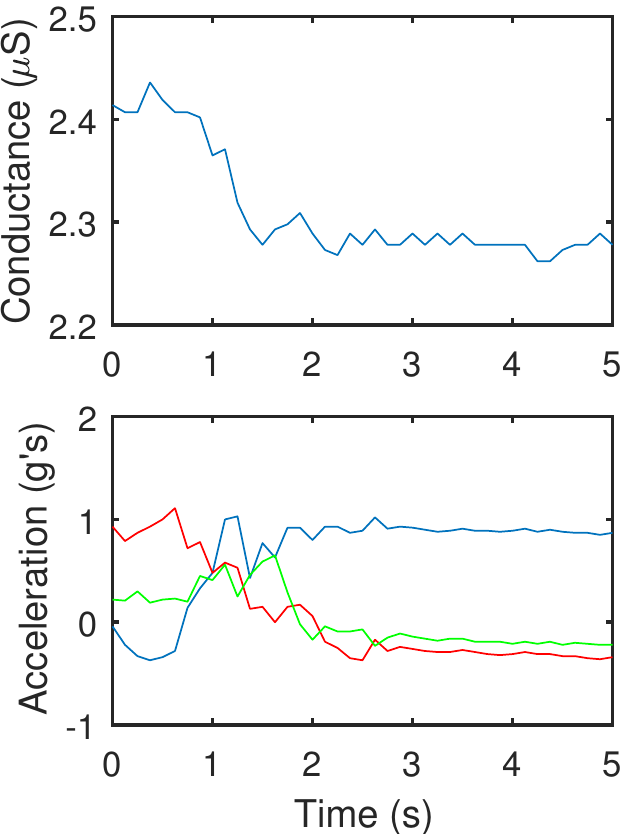}
\caption{Examples of time windows where kNN classifier using all features fails but using EDA only succeeds (left) and vice-versa (right).}
\label{fig:ExampleWindows}       
\end{figure}

In all of our experiments, we found, somewhat surprisingly, that the accelerometer-derived features added {\edit little value to supervised learning algorithms (0.4\% improvement on average). 
This does not necessary imply that the accelerometer data itself has 
little value---it could be that the features we adopted, which are commonly used for activity recognition, are not ideal for MA detection. 
The experts noticed that, on some occasions, even though the acceleration changes, the EDA doesn't get affected at all.} Perhaps such movements do not cause any change in the contact between the EDA electrodes and the skin and thus don't affect the EDA signal. 
{\edit An example of such a time window is shown in the left pane of Figure \ref{fig:ExampleWindows}. 
Conversely, the right pane of Figure \ref{fig:ExampleWindows} shows an example of a time window where the accelerometer data is helpful.} 
Since the accelerometer data {\edit only slightly improve the supervised algorithms}, it is reasonable to expect that they would degrade the unsupervised algorithms, which cannot distinguish between important and irrelevant features without labeled data. This aligns with our experimental findings.
We also observed that the overall performance of the unsupervised learning algorithms can be {\edit very competitive with the supervised ones. We believe this finding has profound consequences, enabling automatic MA detection on a large scale without the need for significant human effort in labeling data!}

One component in a typical data analytics pipeline that we have neglected is feature selection, which is important given that we have constructed up to {\edit 120} features. Adopting a feature selection approach would likely increase the accuracy of each algorithm, but it is unclear how the benefits would be distributed between the supervised and unsupervised algorithms. The evaluation of algorithms for other more complex machine learning settings that lie in between supervised and unsupervised settings, especially semi-supervised learning and transfer learning, would also be of tremendous value for EDA motion artifact detection, {\edit as would evaluation of algorithms such as deep neural networks capable of automatically learning features directly from the raw data.}

%
%
%
%
%


\bibliographystyle{acm-sigchi}
\bibliography{ISWC_2017_EDA_Artifacts}

\begin{thebibliography}{10}

\bibitem{affectivaWhite}
{Liberate yourself from the lab: Q Sensor measures EDA in the wild}.
\newblock White paper, {Affectiva Inc.}, 2012.

\bibitem{Birjandtalab2016}
Birjandtalab, J., Cogan, D., Pouyan, M.~B., and Nourani, M.
\newblock {A non-EEG biosignals dataset for assessment and visualization of
  neurological status}.
\newblock In {\em Proc. IEEE Int. Workshop Signal Process. Sys.} (2016),
  110--114.

\bibitem{boucsein2013electrodermal}
Boucsein, W.
\newblock {\em Electrodermal Activity}.
\newblock The Springer Series in Behavioral Psychophysiology and Medicine.
  Springer US, 2013.

\bibitem{chen2015wavelet}
Chen, W., Jaques, N., Taylor, S., Sano, A., Fedor, S., and Picard, R.~W.
\newblock Wavelet-based motion artifact removal for electrodermal activity.
\newblock In {\em Proc. 37th Annu. Int. Conf. IEEE Eng. Med. Biol. Soc.}
  (2015), 6223--6226.

\bibitem{Dix2017}
Dix, A.
\newblock {Alan Walks Wales data}, 2017.
\newblock \url{http://alanwalks.wales/data/}.

\bibitem{Hammerla2015}
Hammerla, N.~Y., and Pl{\"{o}}tz, T.
\newblock {Let's (not) stick together: Pairwise similarity biases
  cross-validation in activity recognition}.
\newblock In {\em Proc. ACM Int. Jt. Conf. Pervasive Ubiquitous Comput.}
  (2015), 1041--1051.

\bibitem{hernandez2011call}
Hernandez, J., Morris, R., and Picard, R.
\newblock Call center stress recognition with person-specific models.
\newblock {\em Proc. Int. Conf. Affect. Comput. Intell. Interact.\/} (2011),
  125--134.

\bibitem{Hernandez2014}
Hernandez, J., Riobo, I., Rozga, A., Abowd, G.~D., and Picard, R.~W.
\newblock {Using electrodermal activity to recognize ease of engagement in
  children during social interactions}.
\newblock In {\em Proc. ACM Int. Jt. Conf. Pervasive Ubiquitous Comput.}
  (2014), 307--317.

\bibitem{Lara2013}
Lara, {\'{O}}.~D., and Labrador, M.~A.
\newblock {A survey on human activity recognition using wearable sensors}.
\newblock {\em IEEE Communications Surveys {\&} Tutorials 15}, 3 (2013),
  1192--1209.

\bibitem{Natarajan2016}
Natarajan, A., Xu, K.~S., and Eriksson, B.
\newblock {Detecting divisions of the autonomic nervous system using
  wearables}.
\newblock In {\em Proc. 38th Annu. Int. Conf. IEEE Eng. Med. Biol. Soc.}
  (2016), 5761--5764.

\bibitem{EdaExplorer}
Taylor, S., and Jaques, N.
\newblock {EDA Explorer}, 2017.
\newblock \url{http://eda-explorer.media.mit.edu/}.

\bibitem{Taylor2015}
Taylor, S., Jaques, N., Chen, W., Fedor, S., Sano, A., and Picard, R.
\newblock {Automatic identification of artifacts in electrodermal activity
  data}.
\newblock In {\em Proc. 37th Annu. Int. Conf. IEEE Eng. Med. Biol. Soc.}
  (2015), 1934--1937.

\end{thebibliography}
\end{document}